\documentclass[useAMS]{mn2e}
\usepackage{graphicx}

\def\acool{\alpha_{\rm cool}}

\def\hMsol{h^{-1}M_{\odot}}

\def\ergs{\, {\rm erg}\,{\rm s}^{-1}}
\def\erg{\, {\rm erg}}
\def\keV{\, {\rm keV}}
\def\gsim{ \lower .75ex \hbox{$\sim$} \llap{\raise .27ex \hbox{$>$}} }
\def\lsim{ \lower .75ex \hbox{$\sim$} \llap{\raise .27ex \hbox{$<$}} }
\def\simprop{ \lower .75ex \hbox{$\sim$} \llap{\raise .27ex \hbox{$\propto$}} }
\def\apj{ApJ}
\def\mnras{MNRAS}
\def\aap{A\&A}

\begin{document}
 
\title[The flip-side of galaxy formation]{The flip-side of galaxy formation: A combined model of Galaxy Formation and Cluster Heating} 

\author[Bower et al.]
{
\parbox[t]{\textwidth}{
\vspace{-1.0cm}
R. G. Bower$^{1}$,
I. G. McCarthy$^{1}$,
A. J. Benson$^{2}$
}
\vspace*{6pt} \\
$^{1}$Institute for Computational Cosmology, Department of Physics, 
Durham University, South Road, Durham, DH1 3LE, UK. \\
$^{2}$Caltech, MC130-33, 1200 E. California Blvd., Pasadena, CA 91125, USA.\\
\vspace*{-0.5cm}}

\maketitle

\begin{abstract}
Only $\sim$10\% of baryons in the Universe are in the form of stars, yet 
most models of luminous structure formation have concentrated on the 
properties of the luminous stellar matter.  Such models are now 
largely successful at reproducing the observed properties of 
galaxies, including the galaxy luminosity function and the star 
formation history of the universe. In this paper we focus on the 
``flip side'' of galaxy formation and investigate the properties of 
the material that is not presently locked up in galaxies.  This 
``by-product'' of galaxy formation can be observed as an X-ray emitting 
plasma (the intracluster medium, hereafter ICM) in groups and 
clusters. Since much of this material has been processed through 
galaxies, observations of the ICM represent an 
orthogonal set of constraints on galaxy formation models. In this 
paper, we attempt to self-consistently model the formation of 
galaxies and the heating of the ICM.  We set out the challenges for 
such a combined model and demonstrate a possible means of bringing 
the model into line with both sets of constraints.

In this paper, we present a version of the Durham semi-analytic 
galaxy formation model GALFORM that allows us to investigate 
the properties of the ICM. As we would expect on the basis of 
gravitational scaling arguments, the previous model (presented 
in Bower et al.\ 2006) fails to reproduce even the most basic 
observed properties of the ICM.  We present a simple modification 
to the model to allow for heat input into the ICM from the AGN 
``radio mode'' feedback.  This heating acts to expel gas from 
the X-ray luminous central regions of the host halo.  With this 
modification, the model reproduces the observed gas mass 
fractions and luminosity-temperature (L--T) relation of groups 
and clusters.  In contrast to simple ``preheating'' models of the 
ICM, the model predicts mildly positive evolution of the L--T 
relation, particularly at low temperatures. The model is energetically
plausible, but seems to exceed the observed heating rates of
intermediate temperature clusters. Introducing the 
heating process into the model requires changes to a number
of model parameters in order to retain a good match to the 
observed galaxy properties.  With the revised parameters, the 
best fitting luminosity function is comparable to that presented 
in Bower et al.\ (2006).  The new model makes a fundamental step
forward, providing a unified model 
of galaxy and cluster ICM formation. However, the detailed 
comparison with the data is not completely satisfactory, and we 
highlight key areas for improvement.

\end{abstract}

\section{Introduction}

The success of galaxy formation models is usually measured by 
their ability to match key observational properties of the galaxy 
distribution. However, only a small fraction ($\sim 10\%$, Cole 
et al.\ 2001; Lin et al.\ 2003; 
Balogh et al.\ 2001; 2008) of the baryons in the universe end up as luminous 
stars. The vast majority of baryons remain in diffuse form, either because they 
are unable to condense out of the intergalactic medium (for example, because
their host dark matter halos are too small to resist heating from the diffuse
inter-galactic background radiation; see, e.g., Gnedin 2000; Benson et al.\ 2002) or because they are ejected from the star forming regions
of galaxies by strong feedback (White \& Frenk 1991; Benson et al.\ 2003 [Be03]). 
In general, it is difficult to observe 
the intergalactic medium directly: because of its low temperature, its properties must be inferred from metal line studies (e.g., Aguirre et al.\ 2005); however, within groups 
and clusters of galaxies the intergalactic medium becomes sufficiently hot (and dense) that it
can be observed at X-ray wavelengths (Cavaliere et al.\ 1976). This X-ray emitting plasma 
is usually referred to as the intracluster medium (ICM). 

There is a long history of work attempting to explain the observed properties 
of the ICM (e.g., Kaiser 1991; Evrard \& Henry 1991; Tozzi \& Norman 2001; 
Voit et al.\ 2003;  Bode et al.\ 2007; McCarthy et al.\ 2008). 
Generally, it has been concluded that the observed properties cannot be
explained by the gravitational collapse of dark matter haloes alone: the energetics of observed clusters and the scaling of the X-ray emission with system temperature
suggest that an additional heat source is required. For example, gravitational
collapse predicts that the X-ray luminosity of clusters should scale with temperature
as roughly $T^2$ (for $T$ more than a few keV), while the observed relation is much
steeper, scaling as $\sim T^{2.8}$ (e.g., Edge \& Stewart 1991; Markevitch 1998). The steepening of the 
relation can be explained by
heating the ICM so that its central density is lower in lower temperature systems. This
is most efficiently achieved by heating the ICM prior to its collapse so that 
a high minimum adiabat is set, resisting the gravitational compression of the system.
Such preheating models have been explored extensively in the literature 
(e.g.\, Kaiser 1991; Evrard \& Henry 1991; Ponman, Cannon \& Navarro 1999; 
Balogh et al.\ 1999; Borgani et al.\ 2002; McCarthy et al.\ 2002; Muanwong et al.\ 2002).
{Indeed the authors of the present paper have been strong proponents of the 
energetic efficiency of the preheating model. However, the source of the 
preheating energy is rarely explicitly modelled. It is often hypothesised to be 
associated with galaxy formation, or the growth of supermassive black holes,
but there is an inherent tension in these models. The scaling of system entropy with
mass ($K_{\rm vir}\propto T_{\rm vir}\rho_{\rm vir}^{-2/3}\,$\footnote{As is common
in the astronomical literature we indicate entropy by the adiabatic index ($K$) of the
gas rather than the thermodynamically correct logarithmic quantity. $T_{\rm vir}$
and $\rho_{\rm vir}$ are the characteristic temperature and gas density of the system.}) 
makes it difficult to simultaneously preheat the IGM to a sufficiently high 
adiabat that it is able explain the properties of galaxy clusters and yet retain sufficient
low entropy gas in lower mass halo to obtain a realistic galaxy population. This is a 
generic problem --- few models attempt to explain the properties 
of the ICM while simultaneously accounting for the observed properties of galaxies (for
two exceptions see Wu et al.\ 2000 and Scannapieco et al.\ 2001). For example,
while Bower et al.\ (2001) explored the effect of heating during galaxy formation 
on the properties of the ICM, these models did not take into account the back 
reaction of this heating on the formation of galaxies. We briefly explored the a self-consistent
model in Be03, but found that it was not capable of reproducing the 
observed galaxy luminosity function. It is, nevertheless, possible that a successful 
pre-heating model may eventually emerge. Two possible strategies include (1) cooling
a large fraction of the baryons prior to the pre-heating epoch and then slowing the consumption
of this material to prolong star formation to the present epoch, or (2) linking the 
preheating level (at $z\sim 2$) to the mass of the present-day halo. The first scheme
is at odds with the strong feedback required in many current galaxy formation models since gas is
rapidly re-cycled between the cold disk and the halo. The second scheme might be 
effective if entropy excesses are strongly amplified during halo mergers (e.g., Borgani et al.\ 2005). 
However, such a scheme is currently too ill-defined to be implemented in to the semi-analytic models.
While we are currently undertaking a series of numerical experiments to better define
the effect of halo mergers on the entropy distribution of the gas they contain
(McCarthy et al.\ 2007; McCarthy et al.\ in prep), the results are currently difficult
to interpret, in part because of the lack of consistency between SPH and Mesh-code
simulations of galaxy clusters (Voit, Kay \& Bryan 2005; Mitchel et al.\ 2008).
It is also important to stress that the energetic efficiency of the pre-heating model is
only realised if the heating occurs before the gas is incorporated into virialised
haloes. This makes the process intrinsically hard to model in the current semi-analytic
framework.}

In view of the above difficulties, it is useful to take a step back from the problem.
If the properties of galaxies and the low observed stellar mass fraction are set aside,
cooling provides an appealing explanation for the observed scalings of the ICM
(Voit \& Bryan 2001; Muanwong et al.\ 2001). Because the cooling time is closely
related to the adiabat (or entropy) of the gas, lower mass systems (with lower
characteristic entropy) tend to cool out a larger fraction of their ICM. This is 
sufficient to reproduce many of the observed trends in X-ray properties, but the 
implied stellar fractions
are much larger than those observed (for a recent discussion see Balogh et al.\ 2008).
{This suggests that a simpler alternative to the pre-heating model is worth further
investigation: we need to arrange for feedback to eject much of the cooling gas from the 
system before, rather than after, allowing it to form stars. In this paper, we explore such
a model, introducing a self-consistent `radio-mode' gas ejection scheme into the 
Bower et al.\ (2006, hereafter B06)
galaxy formation model. We propagate the ejected gas fractions through the 
merger hierarchy so that the scheme has elements in common with the preheating
scenario discussed above. However, since gas is ejected in 
virialised haloes by `in-situ' heating, it has none of the energetic efficiency of the 
pre-heating scenario and the required energy injection will inevitably be large.}

The over-cooling problem is closely related to the problems of shaping the 
galaxy luminosity function and explaining galaxy ``down-sizing'' and 
the absence of bright blue galaxies at the 
centres of clusters. In B06 we showed that these problems could be 
resolved by including a strong ``radio mode'' of AGN feedback in the models
(see also Croton et al. 2006: we use the term ``radio-mode'' to distinguish 
recurrent, largely mechanical AGN feedback resulting from accretion in 
hydrostatic haloes, from the more radiatively efficient ``quasar-mode''
of AGN activity which we associate with galaxy mergers and disk instabilities). At late times
(low redshifts), massive haloes host galaxies with large black holes so that even a 
small amount of gas cooling out of the ICM and being accreted onto the black hole
results in sufficient energy feedback to offset the cooling. In B06 we assumed that this
set up a self-regulating feedback loop that prevented any significant amount of gas cooling.
Adding the additional requirement that the feedback loop is only effective in hydrostatic
haloes (where the sound crossing time is shorter than the cooling time at the cooling
radius) creates a natural scale at which the efficiency of galaxy formation falls.
This results in a good match to the luminosity function and other observational
constraints on the formation and evolution of galaxies 
(see Birnboim \& Dekel 2003 and Keres et al.\ 2005 for further discussion of 
importance of distinguishing `hydrostatic and ``rapid cooling'' haloes).

In this paper we take the process a step further. We consider the possibility that 
sufficiently massive black holes not only prevent cooling in their host haloes, but
may also inject sufficient energy to expel gas from the halo. As gas is expelled, the 
central density drops, the cooling time becomes longer and the cooling rate, 
and hence energy feedback, become smaller. The system will move to a new lower density 
configuration where the energy feedback just balances the cooling rate. The concept
is appealing since the final configuration is set by the cooling time in the 
halo, while the ejection of gas avoids the excess production of stars. It combines
the simplicity of the scheme suggested by Voit \& Bryan (2001) while offering the
potential to give a good match to observed galaxy properties. The model allows us
to propagate the effects of heating at early epochs to later times, but it does not
implement ``preheating'' in the way envisaged by many previous papers. {As a result, 
the energy requirements of the model we present are larger than in preheating
schemes. Since observational estimates of the pV work required to inflate X-ray cavities 
suggest that jet powers are only comparable to cluster cooling luminosities 
(above 3~keV) this is a significant draw-back (e.g., Birzan et al. 2004; 2008; Dunn\& Fabian 2006;
Best et al. 2007) unless these estimates severely underestimate that total
jet heating power. The model will also struggle
to match the details of the internal properties of clusters --- see the discussion
of the entropy profiles of clusters at intermediate radius in McCarthy et al.\ (2008),
for example. }

Nevertheless, an order of magnitude calculation shows that the model
is worth further consideration. Examining the properties of clusters in the B06 model,
we find that the {\it total} mass of all the black holes in a cluster of mass 
$M=3\times10^{14}\hMsol$ is typically $\sim 5\times10^9\hMsol$. 
{In order to estimate the maximum energy contribution from black hole
growth, we assume the radio-mode dominates 
the growth of these large black holes, and the kinetic power of the jet 
is $0.1\dot{m}_{\rm bh}c^2$} (where $\dot{m}_{\rm bh}$ is the mass growth rate of the black 
hole). Under these assumptions, the total heating energy is $\sim 10^{63}h^{-1}\erg$, while the potential
energy of the baryons is $\sim GM^2f_{\rm b}/r_{\rm vir} = 1.5\times10^{63}h^{-1}\erg$
(where $f_{\rm b}$ is the baryon mass fraction and $r_{\rm vir}$ is the virial radius of the
system).  Since these numbers are comparable, it suggests that black hole heating could 
eject a substantial fraction of the hot gas from the system. At lower halo masses, the black hole
heating would completely dominate the thermal energy of the baryons; while, at higher halo masses, the 
black hole heating becomes a minor perturbation.  Thus, under these assumptions,
the effect of this heating is to establish a new scale of $\sim 3\times10^{14}\hMsol$ on 
which haloes are able to retain their hot X-ray emitting plasma. In the rest of this paper, 
we explore this idea in detail, adding flesh to the order of magnitude calculation outlined above. 
In particular, we take full account of the different channels for black hole mass 
growth (we assume that black hole growth occurs through the QSO mode does not provide
heat the ICM efficiently) and for the effect of heating in subhaloes that are subsequently 
accreted by the growing cluster.

The methods we adopt here are semi-analytic and based on the techniques
described in detail in Cole et al.\ (2000) (see the recent review by Baugh 2006). 
We are able to implement feedback on a macroscopic scale without attempting to resolve the detailed 
physical processes
that heat and eject the ICM.  Much simulation work is being devoted to studying the 
formation of jets and their interaction with the surrounding ICM 
(Quilis et al.\ 2001; Churazov et al.\ 2001; Dalla Vecchia et al.\ 2004; Heinz et al.\ 2006; 
Sijacki \& Springel 2006). Ultimately, these mechanisms need to
be incorporated into cosmological scale simulations of galaxy formation and the
ICM (Sijacki et al.\ 2007; Okamoto et al.\ 2008).  Unfortunately, the myriad
of important physical processes make this direct approach extremely difficult and computationally 
expensive.  Semi-analytic methods, such as those adopted in the present study, allow a wide range of 
possible physical processes to be explored with a greatly reduced computational effort.
Ultimately, however, the details of the processes we model will need to be justified by high resolution
numerical simulations and observations. 

The approach we present here is intended to capture the broad-brush energetics and 
integrated properties of clusters. We assume that the heating effect of the AGN can 
be captured by a single number that measures the non-gravitational heat input into 
the system and we adopt a particular form the modification of the system's density profile.
In reality the situation is likely considerably more complex: for example, the radial 
dependence of the heat input may vary between systems (e.g., Heinz et al.\ 2006 have argued that 
halo mergers play a vital role in mixing the heat input from jets into ICM),
or the energy in an infalling subsystem might be distributed in different ways
depending on the shocks generated as it falls into the main halo (McCarthy et al.\ 2007).
As a result of these processes, systems may have different density profiles even though
the total energy input is the same and we cannot expect to recover the detailed radial structure of 
clusters. However, despite this simplification, we 
will see that the model already captures the global features of observational 
data well, including the scatter in the X-ray luminosity--temperature (hereafter L--T) correlation. 

The structure of the paper is as follows. In Section~2, we describe how we implement the
``radio mode'' heating of the ICM. This section includes a discussion of the
parameter values that are modified from those in B06. We present results from 
the model in \S3, initially focusing on the X-ray luminosity--temperature (L--T) correlation, 
and then moving to gas mass fractions and galaxy properties.
 We present our conclusions and discuss how
the model can be developed in \S4. Through out we use units of $\hMsol$ for masses. 
For comparison with X-ray observations, however, the choice of H$_0$
does not scale out of the relations, and we adopt 
$H_0=73$ $\hbox{km}\,\hbox{s}^{-1}\,\hbox{Mpc}^{-1}$. 
The model assumes $\Omega_b = 0.045$, $\Omega_M=0.25$ and 
$\Omega_{\Lambda}=0.75$.

\section{Modeling the heating of the ICM}

\subsection{Semi-analytic implementation}

In this paper we introduce a relatively simple modification to the B06 model in 
order to take into account the heating effect of the AGN.  The basis of the 
method is to compute the feedback energy from the AGN as a function of the 
cooling rate. In hydrostatic haloes, the feedback energy is used to
redistribute or eject the gas from the halo, thus reducing the system's central 
density. We make this modification by reducing the density normalisation
while keeping the shape of the profile unchanged. The results are not strongly dependent
on the details of where the ``ejected'' gas is placed so long as it is removed from
the X-ray luminous central regions of the cluster. 
Because there is only one parameter determining the modification (specifically, the
ratio of cumulative energy injected by the AGN to the thermal energy of the
halo --- see below) we can track the change in the profile by accumulating the
additional energy that is input into the profile (Wu et al.\ 2000; Bower et al.\ 2001).

In subsequent time steps, a new cooling rate and feedback energy is calculated.
If the feedback energy still exceeds the radiated energy, the gas distribution is further 
adjusted. This process continues until the system reaches a stable configuration 
where the cooling rate is balanced by the heating rate. 

We compute the heating power ($L_{\rm heat}$) available from the AGN as the smaller of
$$
   \epsilon_{\rm SMBH} L_{\rm Eddington} 
$$
and
$$
   \eta_{\rm SMBH} 0.1 \dot{M}_{\rm cool} c^2
$$
where  $L_{\rm Eddington}$ is the Eddington luminosity of the black hole and
$\dot{M}_{\rm cool}$ is the cooling rate of the halo (in the absence of radio-mode
feedback). $\epsilon_{\rm SMBH}$ and $\eta_{\rm SMBH}$ are parameters controlling
the disk structure and the efficiency with which cooling material can be accreted by the black hole.
The first condition corresponds to the Eddington luminosity criterion used
in B06. {At first sight, it might seem that the Eddington luminosity is 
not a relevant criterion for radio-mode feedback. However, we emphasise that
the limit we impose relates to the accretion disk structure, rather than
the maximum feedback energy itself. Efficient jet production is thought to 
be associated with a geometrically thick, advection dominated disks (e.g., Rees et al.\ 1982;
Meier 2001; Churazov et al.\ 2005). 
If the accretion rate is too high, current models suggest that the vertical height 
of the disk will collapse with a corresponding drop in jet efficiency.
Esin et al.\ (1997) suggest that this structural change occurs at $\dot{M_{\rm BH}} \sim 
\alpha^2\dot{M_{\rm Ed}}$ (where $\dot{M_{\rm BH}}$ is the accretion rate on to the black hole,
$\dot{M_{\rm Ed}}$ is the Eddington accretion rate and $\alpha$ is the disk viscosity parameter).}
We adopt an disk structure parameter of $\epsilon_{\rm SMBH}=0.02$.\footnote{
Note due to error in B06, cooling luminosities were over estimated by a factor
$4\pi$. Thus, while the paper quotes the efficiency parameter $\epsilon_{\rm SMBH}$ 
as 0.5, this should have been $0.5/4\pi = 0.04$. With this correction the rest of the 
parameters and results are unchanged.} This is in broad agreement with plausible
accretion disk viscosities (e.g., McKinney \& Gammie 2004; Hirose et al. 
2004; Hawley \& Krolik 2006). If the accretion rate becomes higher, the 
efficiency actually drops, as the accretion disk becomes thinner and 
its magnetic field threads the plunging region around the black hole less
effectively. In this case, much more of the accretion disk energy is 
radiated and is not available for ``radio mode'' feedback. 

{ The second criterion corresponds to the accretion power released when the 
cooling gas reaches the black hole. We assume that a maximum fraction, $\eta_{\rm SMBH}$, 
of the cooling gas is available to power feedback from the black hole. 
This approach differs from B06 in that it is not only the Eddington luminosity of 
the black hole that limits the available feedback but also the amount of material cooling 
out of the halo. Before significant material has been ejected from the halo, 
the first criterion usually limits the energy output. We adopt an efficiency of
$\eta_{\rm SMBH}=0.01$ (i.e., only 1\% of $\dot{M}_{\rm cool}$ reaches the black hole)
and assume that this mass accretion results in a jet power output of $0.1c^2\dot{M}_{\rm BH}$
(i.e., 10\% of the mass accretion rate). The latter is easily compatible with the efficiency 
of jets expected from advection 
dominated accretion disks around spinning black holes (e.g., Meier 1999; 2001; Nemmen et al.\ 2007).
Allen et al.\ (2006) compare the jet power (measured from the cavity PdV work) with the 
Bondi accretion power of several nearby X-ray luminous ellitpical galaxies. They find 
that the jet power is 2\% of the rest mass energy accretion rate at the Bondi radius.
This corresponds to $\eta_{SMBH}=0.01$ if we equate the Bondi accretion rate with 5\% of
$\dot{M}_{\rm cool}$. In practise, however, it is more relevant to compare the heating
power with the system's cooling luminosity. We find that our systems have a bimodal
distribution of energy injection rates, with many of the systems in a passive phase.
Amongst the active systems, however, we find that the injected power typically
exceeds the total cooling rate by a factor 10--100. Thus the model significantly exceeds the
energy rates estimated on the basis of cavity PdV work (e.g., Fabian et al 2003; 
Birzan et al 2004; Dunn et al 2005). As Nusser et al.\ (2006) and Best et al.\ (2007)
emphasise it is likely that this underestimates the true heating rate, but probably
not by as large a factor as required by the model.  We return to this point
in \S4.
Finally, we note that the energy feedback that this model requires from the radio mode implies
additional black hole mas growth. We compute the radio-mode contribution to the mass growth rate
of the black hole as $\dot{M}_{\rm BH} = L_{\rm heat}/0.1c^2$. Combining this with
the limit on the heating rate implies that $\dot{M}_{\rm BH} < \eta_{\rm SMBH} \dot{M}_{\rm cool}$.
}

The heating energy given by the above criteria is then compared to the 
cooling luminosity of the system.
If the heating energy is greater than the cooling luminosity,
the excess energy ejects mass from the X-ray emitting region of the halo.
$$
   {dM_g\over dt} = {{L_{\rm heat} - L_{\rm cool}}\over {1\over2} v_{\rm halo}^2}
$$
where $v_{\rm halo}$ is the circular velocity of the halo at the virial
radius.  The divisor provides an estimate of the energy required 
to eject the gas from the X-ray emitting region.
We limit the amount ejected in any one time step to $<$50\% of the current hot gas 
mass content in order to ensure the numerical stability of the code. 
{It is important to note that the gas ejected in a time step greatly exceeds 
the amount of gas that tries to cool out of the halo. In this respect our model
differs significantly from ``circulation flow'' models (e.g., Mathews et al.\ 2003;
McCarthy et al.\ 2008) in which gas is heated by the AGN as it cools out of the 
hot phase.}

As the AGN ejects material, the halo profile is rescaled in density to match the new hot gas mass.
We do not alter the core or slope of the gas density profile. This is in broad agreement 
with current observations of groups and clusters and results in important 
simplifications of the cooling calculations that allow us to maintain a high computational
speed for each halo. At the next time step, $L_{\rm cool}$ will be reduced.
Of course this choice of how to modify the halo profile is somewhat arbitrary.
Ideally, we might consider the change in entropy of the heated material
and propagate this forward through the hierarchy, re-deriving the modified
halo profile from its hydrostatic equilibrium in the gravitational potential 
at each step.  This is not possible with our current code because we cannot 
yet propagate entropy modification through the merger hierarchy (see McCarthy et al.\ 
2007 for recent progress). However, the main impact of the scheme we consider is to
alter the core density of the gas, and thus to modify the central cooling time
and the X-ray luminosity of the system. 
{An important point to note here is that the global properties of the X-ray emitting
plasma are dominated by the gas within the central regions of the halo. Thus,
global properties (such as X-ray luminosity and X-ray emission weighted
temperature) are insensitive to how the gas that is removed from the core
is distributed outside the core, and we would expect to obtain similar results
whether the heated gas is ejected completely from the system or whether the outer 
radial profile is modified so that the gas is held at relatively low density in 
the outer parts of the system. Nevertheless, there is a clear need to address
this issue in future models: the radial distribution of system entropy and
temperature provide sensitive probes of the gas distribution that will allow
us to observationally constrain the manner in which gas is re-distributed or ejected
within the cluster (e.g., McCarthy et al.\ 2008). 
The present model is too simplistic to rise to this challenge.
}
Another point is that because we specify the effect in terms of a change 
in the density profile, we are not requiring that 
it is the lowest entropy material that is ejected from the system. Indeed, 
the ejected material is drawn from a wide range of entropies and the entropy
of material at the virial radius rises as the density normalisation is reduced.
One can imagine that there may be more energetically efficient ways in which the
X-ray luminosity of the system can be reduced by adding thermal energy to 
the gas. However, this uncertainty combines directly with the uncertainty in the
efficiency of the black hole accretion.

{The strategy we adopt here differs significantly from preheating approaches. 
In preheating scenarios, gas is heated to high temperatures (or, more correctly, entropies) 
in low mass haloes preventing it from following the collapse of the dark matter 
hierarchy. In the present model, the ejected hot gas is gradually recaptured as 
groups merge to form more massive systems and gravitational potential deepens.}
We treat this by allowing a fraction of the ``ejected'' gas to be re-integrated
into the hot gas component after the merger. The ejected gas mass in the
new halo is given by 
$$M_{\rm ejected, new} = 
       M_{\rm ejected, progenitor}\left(1 - {M_{\rm halo, progenitor}\over M_{\rm halo, new}}\right)$$
which is consistent with energy conservation.  This component of the
model is important because it allows massive systems to retain a much
larger fraction of baryons in their hot component than in the smaller
systems. This is crucial in matching the observed scalings of X-ray properties.

In our current model, we assume that the gas distribution is a $\beta$-profile
(Cavaliere \& Fusco-Femiano 1976)
with $\beta=2/3$, and that the gas is isothermal. This maintains compatibility with B06
and results in some key simplifications of the energy calculations that allow us to use 
look-up tables to evaluate the energy integrals and hence maintain a high computational
speed. The code speed is essential since we will later need to evaluate many possible
parameter sets in order to identify the best fitting models.

{The details of the treatment of gas cooling are described in Cole et al.\ (2000)
and Be03, but it is helpful to summarise them here. A fundamental
building block of the code is the idea that a new halo is created when its mass
of its main progenitor doubles. At this point, the density normalisation of the 
halo is set from the hot gas mass, and this time step is used the reference point 
for the age of the halo. As the halo ages, we compute the cooling radius as a function
of the age of the halo. The amount of gas considered for cooling in any given time step is 
computed as the difference between the mass cooled in the previous time step and
the mass within the current cooling radius. Whether this material successfully cools
or not is dependent on the AGN feedback that we have described above. The present 
code develops the approach in Be03, so that the density normalisation 
increases in response to hot gas accreted by the system (both from mergers and from diffuse 
accretion, see Be03) and decreases in response to the ejection of material by the AGN.
We compute the luminosity of the hot halo using the density normalisation
defined above and metallicity-dependent Sutherland \& Dopita (1993) cooling tables
using the metal abundances self-consistently calculated by GALFORM.
Note that this calculation implicitly assumes that as material cools out of the
centre of the halo it is replaced by material from larger radius: in line with
GALFORM's Lagrangian approach to the cooling calculation, we do not adjust the overall
density normalisation as material cools out until the halo has doubled in mass.
In practise these subtleties have only a weak impact on the predicted X-ray luminosity
since, by construction, little material cools out in hydrostatic haloes with effective AGN.
}

The $\beta$-model has one adjustable parameter --- the gas core radius. 
We set the core radius of the gas distribution in order match the luminosity
of the highest mass clusters. In practise, the luminosity of these systems
depends little on the energy used to eject gas from the system because the
hot gas fraction is always high. The effect of increasing the energy
that is injected by the AGN is to increase the fraction of hot gas that is
``ejected'' from lower mass groups. We find that a core radius of 0.025 of the 
virial radius gives reasonable X-ray luminosities for massive systems
(where we expect the effect of AGN heating to be small).
This approach is preferable to directly adopting an
observed core radius since it reduces our dependence on the exact shape of the 
radial density profile. We note that the core radius we
adopt is larger than that typically found in radiative 
cosmological simulations\footnote{But note that such simulations typically 
suffer from the over-cooling problem, possibly because they often neglect the 
heat input from AGN.  Introducing a more efficient form of feedback into such 
simulations, such as AGN feedback, could result in larger core radii.}, but is 
comparable to those found in simulations
that do not include cooling (e.g., Frenk et al.\ 1999; Voit, Kay \& Bryan 2005). 
The physical origin of this baseline profile
is beyond the scope of the model we present here: although further work
is clearly justified to remove this limitation of the current model.
As well as setting the normalisation of the bright end of the L--T relation,
the choice of core radius also plays an important role in determining the 
amount of gas that must be ejected from the system before the cooling rate
drops below the critical threshold for feedback.

\subsection{Revised model parameters}

By introducing this form of heating into the model, we make significant changes to the way
in which galaxies form, and we must adapt the model parameters to have any chance
of obtaining a model that agrees with the observational data (both the X-ray and galaxy properties). 
We start by re-running the Millennium simulation models described in B06
using the revised code. This use halo trees extracted from the Millennium
Simulation (Springel et al.\ 2005, Lemson et al.\ 2006) to provide the merger history of
dark haloes within which we compute the galaxy and ICM properties. The new physics
that we have introduced has a significant impact on galaxy propeties. As a result of the 
AGN heat input, haloes are more likely to be hydrostatic in the new model, and 
(if we make no adjustment to the parameters) we find that the break in the luminosity is too faint
to be compatible with observations. We therefore search for a revised set of
parameters that provides a better description of galaxy formation in the new model.
The final parameter values result in a reasonable match to the observed galaxy luminosity 
function (see \S3.3). In terms of the X-ray relations, however, the results 
are insensitive to the choice of most of the parameters described below. 

\begin{table}
\begin{center}
\begin{tabular}{lcc}
Parameter&  B06&   new model\\
\hline
\noalign{\smallskip}
stellar yield    &    0.02&  0.04\\
$\epsilon_{\rm SMBH}$&  0.04&   0.02\\
$\eta_{\rm SMBH}$&  --&   0.01\\
$\alpha_{\rm cool}$&    0.58&  0.6\\     
$\alpha_{\rm reheat}$&  0.92&  1.0\\     
$v_{\rm hot}$&             485&   400\\
$\tau_{\rm 0,star}$&             350&   400\\
$\tau_{\rm 0,mrg}$&              1.5&   2.5\\
\hline
\end{tabular}
\caption{Comparison of parameter values in the B06 and the new model presented
in this paper. The effect of the parameter changes is described in the text.}
\label{tab:param_table}
\end{center}
\end{table}

The revised parameter values are listed in Table~\ref{tab:param_table}. The values 
used in B06 are given for comparison. It is worth briefly outlining the rationale for
the changes that we have made. 
\begin{itemize}
\item  We doubled the stellar yield in order to obtain a better match to the stellar 
colours (see discussion in Font et al, 2008) and to increase the ICM metal abundances.  
With this higher yield, the ICM metal abundance works out well 
compared to observational data in cluster cores (e.g., De Grandi et al.\ 2004). 
On average we find $0.4 Z_{\odot}$ (where we use the Grevesse \& Sauval (1998) measurement 
of the solar abundance) with a scatter of $0.1 Z_{\odot}$.
However, we find no significant dependence on the halo mass which appears to conflict with 
recent results for galaxy groups (e.g., Rasmussen \& Ponman 2007).  We caution, however, a rigorous 
comparison to the observational data is difficult because inhomogeneities are weighted by luminosity and the 
observations must be averaged over the observed abundance gradients. Moreover, the model 
presented here uses the instantaneous recycling approximation and does not distinguish 
between SN type Ia and type II products.

\item $\epsilon_{\rm SMBH}$, $\eta_{\rm SMBH}$.  
{These parameters control the maximum rate of black hole accretion
in the radio mode (see \S2.1).  Smaller values reduce the rate at which haloes
can eject mass and thus results in more scatter in X-ray luminosity lower temperature systems 
but reduced scatter in the high mass systems. Larger values of the parameter also result in 
greater departure from linearity in the relation between galaxy bulge mass and black 
hole mass. The parameters have been set to provide a good match to the L-T relation while 
preserving uniformly high gas fractions in the most massive clusters.}

\item  $\tau_{\rm 0,mrg}$. We find that we need to increase the merger timescale relative to the 
values used in B06, and relative to the stripping calculations of Benson et al.\ (2002). 
Leaving this parameter fixed at its fiducial value, we find that the model
generates a tail of bright galaxies and the bright end of the luminosity function
does not drop away sufficiently rapidly to match observations.

\item  $\alpha_{\rm cool}$, $\alpha_{\rm reheat}$, $v_{\rm hot}$, $\tau_{\rm 0,star}$. 
These four parameters adjust the location of the break in the luminosity function and 
improve the fit of the final model to the observed galaxy luminosity function. 
$\acool$ determines the ratio of free-fall and cooling times at which haloes
are taken to be hydrostatic (as opposed to being classified as ``rapid cooling''), 
so that only when $t_{\rm cool}(r_{\rm cool}) > \alpha_{\rm cool}^{-1} t_{\rm 
ff}(r_{\rm cool})$ is the AGN feedback effective\footnote{Equation (2) of B06 is incorrect;  
$\alpha_{\rm cool}$ should be replaced by $\alpha_{\rm 
cool}^{-1}$.}. The value of this parameter primarily affects 
the location of the luminosity function break. Increasing $\alpha_{\rm cool}$ makes 
the characteristic luminosity of galaxies, $L_*$, fainter. The remaining
parameters primarily act to moderate the normalisation of the luminosity function.
In outline,  $\alpha_{\rm reheat}$ determines the timescale on which gas ejected by supernova
winds becomes available for cooling; $v_{\rm hot}$ controls the efficiency of supernova
feedback (we apply the same values to quiescent star formation and star formation in
bursts); $\tau_{\rm 0,star}$ set the timescale for star formation. Further details
and the definition of these parameters are  given in B06.
\end{itemize}

Although it seems that the model parameters differ only slightly from the values used in B06, 
this hides a chain of dependencies. The effect of increasing the
stellar yield is to shift the division between hydrostatic and rapid cooling to higher
mass haloes. If left uncompensated for, this makes the break of the luminosity function brighter.
In order to maintain a good fit, the value of $\acool$ needs to be increased. This is 
seen in the ram pressure stripping model of Font et al.\ (2008). However, the additional
physics that we have introduced to model the heating of the ICM compensates for
this effect of the increased yield, almost exactly restoring the luminosity function
break to its original position.  If we had not introduced the higher yield, a good fit
to the luminosity function would have required a smaller value of $\acool$. 
The final fit to the luminosity function and discussion of other galaxy properties is presented in \S3.3.

\section{Results}

\subsection{The L--T relation}

\begin{figure}
  \includegraphics[width=8.6cm]{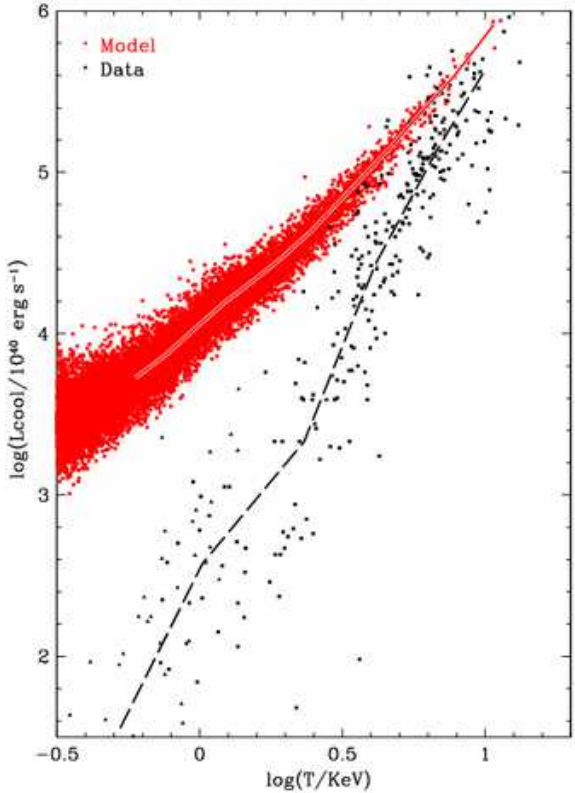}
\caption{The bolumetric Luminosity--Temperature relation for the model in the absence 
of AGN heating. The model parameters are based on B06 with the adjustments
described in the text. Black squares are data from Horner (2001), based
on {\it ASCA} measurements, triangles are from Osmond \& Ponman (2004). 
Red points are the 
predicted X-ray luminosities of model haloes. The temperature plotted for 
the models has been corrected for the systematic offset between the spectral 
temperature and the halo virial temperature.
Red and black lines show the median relations for models and data respectively.
As expected from gravitational scaling, the model relation is too shallow 
compared to the data. 
} 
\label{fig:LT_no_heating}
\end{figure}

The relation between bolometric X-ray luminosity and system temperature is a basic
observational correlation that can readily be measured over a wide range of system
temperatures. We first compare the data with the model in the absence of any heating. 

We use the large data sample compiled by Horner (2001) as the basis of our 
comparison. The data spans a wide range of system temperatures and is relatively 
unbiased with respect to system surface brightness. This is preferable to using 
data on smaller less homogeneous samples, which are often 
preferentially picked to have high surface brightness.  Horner (2001) derived bolometric luminosities and mean system temperatures (both uncorrected for the presence of ``cool cores'') from {\it ASCA} data.
We supplement this with group data from the GEMS project (Osmond \& Ponman 2004),
converted to bolumetric X-ray luminosity assuming a temperature of 1~keV. For the model comparison, 
we compute the bolometric
X-ray luminosity using the Sutherland \& Dopita (1993) cooling tables, assuming the
self-consistent metal abundance computed by GALFORM.   The
temperature most readily derived from the models is the virial temperature 
of the halo. However this is systematically different
from the X-ray spectral temperature ($T_{\rm spec}$) that is computed from the data. We estimate
this correction using a hydrostatic model to determine the radial temperature
distribution and then appropriately weight the radial contribution to the
X-ray luminosity from each shell as described in Mazzotta et al.\ (2004). 
This correction increases the temperature plotted for each halo by typically 10\%
compared to the virial temperature, but is 10\% less than the emission weighted
temperature of the system.

Fig.~\ref{fig:LT_no_heating} shows the L--T plot for the model in the absence of
AGN heat input.  Adopting a fiducial gas core, $r_{\rm core} =0.025 r_{\rm vir}$, the 
model matches fairly well for the most massive clusters (that is, when the model
virial temperatures corrected to X-ray spectral temperatures), 
but the slope of the model relation is clearly far, far too shallow.  Black and red solid lines show the
median relations for the model and data respectively. This discrepancy is expected
--- it is well known that the scaling expected from gravitational collapse
is unable to explain the observed L--T relation (e.g., Kaiser 1991; Henry \& Evrard 1991): 
the expected system luminosity scales as the mass of the system times the relatively 
weak temperature dependence of the cooling function. An improved match requires that
we add energy to the lower mass systems so that their central densities are lower
and the model relation becomes steeper. Note that heating these systems tends to 
affect their temperatures only weakly, because the system temperature is determined
by the depth of the gravitational potential rather than the specific energy of
the gas (Voit et al.\ 2003). As we will see, by including the effects of the AGN heating,
we can obtain a much better match to the observed L--T relation. Although this does
affect the properties of the galaxies that form, the heating can be compensated by 
adjusting the parameters controlling the luminosity function.

\begin{figure}
  \includegraphics[width=8.6cm]{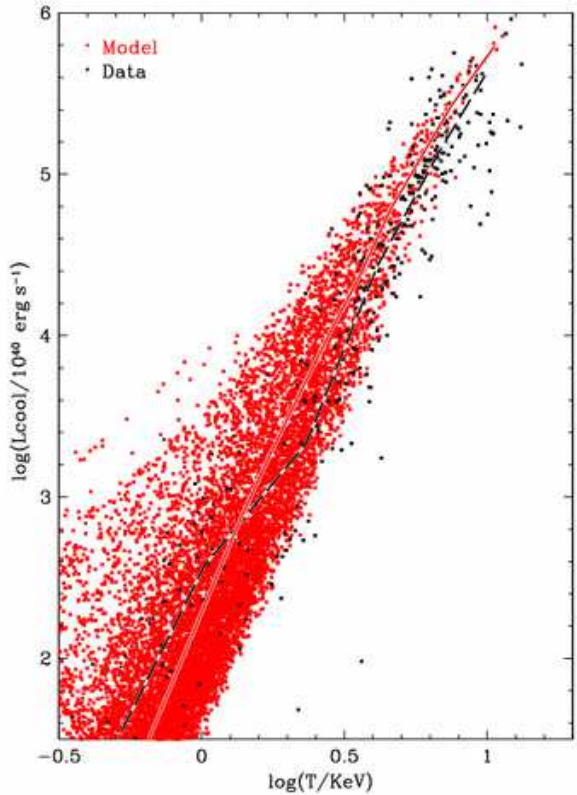}
\caption{The L--T relation for the model when AGN heat input is taken into
account. Points and lines are described in the caption to Fig.~\ref{fig:LT_no_heating}. The
effect of the AGN heating is to steepen the relation by preferentially
ejecting gas from lower mass systems.}
\label{fig:LT_heating}
\end{figure}

We now show the effect of including AGN heating in the model.
The effect of the gas ejection is to reduce the predicted luminosities of
lower mass systems. This occurs because the cooling time of the lower mass 
systems is shorter and thus they initially supply more material to the 
the AGN resulting in larger feedback energy (per unit gas mass). As we have
described above, hot gas is then ejected from the X-ray emitting region 
until the cooling rate (and the X-ray luminosity) of the system drops below a critical value. 
In contrast, the most massive systems have such long cooling times that little 
material is able to cool and they therefore retain close to the universal 
baryon fraction.

The model L--T relation is shown in Fig.~\ref{fig:LT_heating}. The model
relation is now significantly steeper and in much better agreement with the observational
data. The model and observed median relations have similar slope and 
(when the model temperatures are corrected to $T_{\rm spec}$) normalisation. In addition
to matching the median slope of the observed data, the model also shows a similar 
variation in the scatter along the relation. In particular, below a temperature of $\sim 3$ keV, 
the model points fan out to fill a triangular region of the L--T plot. The success of the 
model is evident in comparison with Fig.~\ref{fig:LT_no_heating}. 

The diversity of model groups in the low temperature region 
arises from the range of merger histories with data points lying towards
the high luminosity edge having recently undergone rapid mass growth. During the rapid growth
phase, gas mass is added to the system and the importance of the past heat input 
decreases relative to the gravitational potential of the new halo. Over time,
the AGN injects further energy in order to re-establish an equilibrium state
and the system will move towards the main relation.
Groups on the low luminosity side correspond to systems with unusually slow mass 
growth rates. This is a key success of the model --- reproducing the diversity of the 
X-ray properties of groups is something that must be added ``by hand'' to
preheating models (see discussion in McCarthy et al.\ 2007). However, although the
general shape of the distribution matches reasonably well, 
there are rather too many low-mass systems with luminosities $\sim 10^{43} \ergs$.
These are systems in which AGN feedback is active but has yet to sufficiently
suppress the existence of the hot halo. Possibly this results from making a sharp distinction
between hydrostatic and rapid cooling haloes in the model. In reality we might expect
the AGN feedback to be partially effective in systems close to this division.  This is
an effect that will need to be calibrated against realistic numerical simulations
or by comparison with the observed scatter in massive clusters.

\subsection{Gas Mass Fractions}

\begin{figure}
  \includegraphics[height=8.4cm, angle=270]{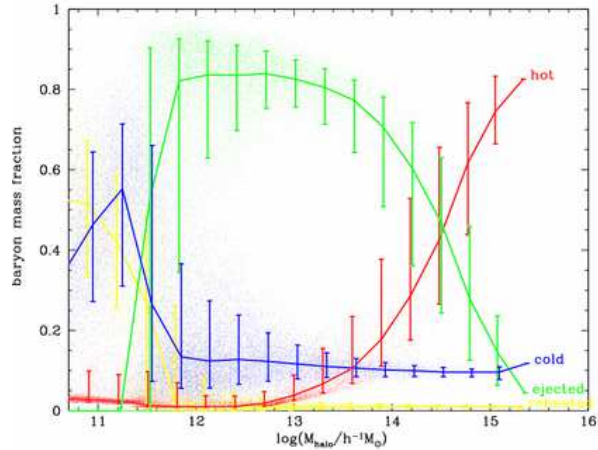}
\caption{The variation in baryon mass content as a function of halo
mass. For a random sample of haloes, red points show the hot, X-ray emitting gas mass; 
green points the ``ejected'' gas mass; the
total of ``cold'' baryons (cold gas within galaxies and stars) is shown
in blue. Yellow points show the ``reheated'' material that has been expelled
from galactic disks but has not yet been incorporated into the haloes hot gas component.
The correspondingly coloured lines show the median baryon fraction in each component,
with error bars showing the 10 to 90 percentile range. The transition from rapid
cooling to hydrostatic haloes occurs at $M_{\rm halo}\sim 10^{11.5}\hMsol$ and results in 
a drop in the cold and reheated components and a rapid rise in the fraction of
baryons ``ejected'' from the halo by the AGN feedback. Above $\sim10^{14}\hMsol$,
AGN feedback becomes less effective and the haloes gain a substantial X-ray emitting
halo.}
\label{fig:fbaryon}
\end{figure}

As we have outlined in the previous section, the model's successful match to the
observed X-ray luminosities of groups and clusters is achieved by ``ejecting''
a large fraction of the hot baryons from the X-ray emitting regions of the lower
temperature haloes. It is interesting to explore this dependence in more detail.
Fig.~\ref{fig:fbaryon} shows the variation of the baryon fraction (expressed
a mass ratio relative to the total baryonic mass of the halo)
as a function of halo mass. Different colours show the mass fractions for the hot gas
(red); the ``ejected'' mass (green) and the cooled gas (i.e., the sum
of cold gas within galaxies and stars).  The ``reheated'' gas (that has
been expelled from galaxies' gas disks by supernova driven feedback,
but has not yet been incorporated into the
hot halo --- see B06) is shown in yellow.  This figure 
illustrates the existence of a characteristic mass scale above which
most of the baryons are in the hot gas phase ($10^{14.5} \hMsol$). 
Below this mass, an increasing fraction of the gas mass is likely to have 
been ejected from the system. However, there is considerable scatter in 
the actual X-ray emitting gas mass, which leads to the scatter in the L--T relation. 

\begin{figure}
  \includegraphics[height=8.4cm, angle=270]{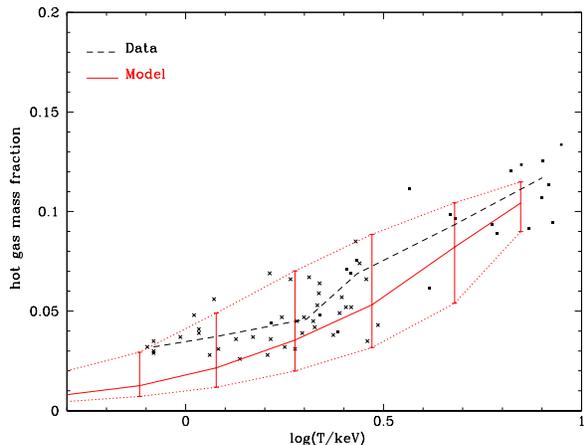}
\caption{The variation in hot gas mass fraction as a function of spectroscopic
temperature. The red line shows the median X-ray emitting gas mass fraction 
of the model haloes.
The scatter in model is show by the error bars and dotted lines, which 
show the 10 to 90 percentile range in each temperature bin.
Measurements of the hot gas mass fraction are shown as solid black points, 
while the dashed black line shows the median fit to the observational data.}
\label{fig:fgas}
\end{figure}

In figure~\ref{fig:fgas} we focus on the hot gas mass fraction of the 
higher mass systems, where we can compare the model gas fractions 
with observational data. In order to do this, we
show the ratio of the hot gas mass to the total mass of the system as a function
of the system's X-ray temperature. {The data (solid black points) are taken 
from profile measurements for clusters from Vikhlinin et al.\ (2006) and 
Pratt et al.\ (2006), and presented at an overdensity of 2500 
(see McCarthy, Bower, \& Balogh 2007). These are supplemented by observations of 
galaxy groups taken from Sun et al.\ (2008), again within an overdensity of 2500.
We compare these measurements with the prediction of the model at the 
same overdensity (taking into account the different profiles of gas an dark matter,
the gas fraction within an overdensity of 2500 is 0.766 of that of the whole
cluster). This region accounts for more than 70\% of the X-ray 
luminosity of the system and thus the gas fraction is closely tied to
the success of the model in accounting for the observed L--T relation. 
A comparison of the gas fractions at larger radius requires us to accurately
specify what happens to the ``ejected'' gas mass: specifically we must decide
whether this material is completely ejected from the halo or stored at large
radius as a small deviation from the beta profile. Comparison with the data
at larger radius is also more difficult and fraught with sample biases.

The median dependence in the model is shown
as a solid red line with error bars showing the scatter.
The model agrees reasonably well, and the trend of rapidly rising gas fraction 
around $T_{\rm spec}\sim3\keV$ is seen in both the data an the model. At lower
temperatures, the data tend to suggest somewhat higher gas fractions than
predicted by the model. However, the data shown here is probably biased to the 
most X-ray luminous galaxy groups. 
The plot also emphasises the large scatter in the hot gas mass fractions
of haloes at a given mass. The scatter in the model appears quite comparable to
that seen in the data. It will be intriguing to see if this comparison holds
up as the sample sizes increase and the sample selection becomes more 
representative. 

Finally, we note that if the comparison is made at
larger radius (lower overdensity), both the data and models predict
higher gas fractions. However, the group data become more
discrepant with the model --- this is in part due to the exclusion of the
lowest surface brightness groups, but may also indicate that the ``ejected''
gas in the model tends to accumulate as a modification of the beta-profile
at large radius in real-world systems (cf., Arnaud \& Evrard 1999; 
Sanderson et al.\ 2003). Detailed 
comparison of entropy and density profiles is an avenue that we will 
explore in future papers.
}

Returning to Fig.~\ref{fig:fbaryon} it is interesting to examine the role of 
gas ejection and feedback in lower mass haloes. Below a halo
mass of $\sim 10^{14} \hMsol$, the X-ray emitting mass fraction is low, creating
the steep slope of the L--T relation seen in Fig.~\ref{fig:LT_heating}.
Without AGN heating, these haloes would have substantial hot gas haloes resulting
in excessively high X-ray luminosities.
Below $\sim 10^{11.5} \hMsol$, haloes are no longer hydrostatic, and the
AGN feedback is assumed to become ineffective. There is a rapid transition
to a regime in which the mass fraction is dominated by stars, cold gas 
and ``reheated'' material (which has been ejected from the galaxy's cold gas disk 
by supernova feedback).  Galaxies in this regime have little hot halo
gas: since the halo cooling time is very short, the reheated phase becomes
dominant. Galaxies in this phase are dominated by a galactic fountain --- material
is expelled from the disk, falling back to the disk on a dynamical timescale.

\subsection{Evolution of the L--T relation}

\begin{figure}
\includegraphics[width=8.6cm]{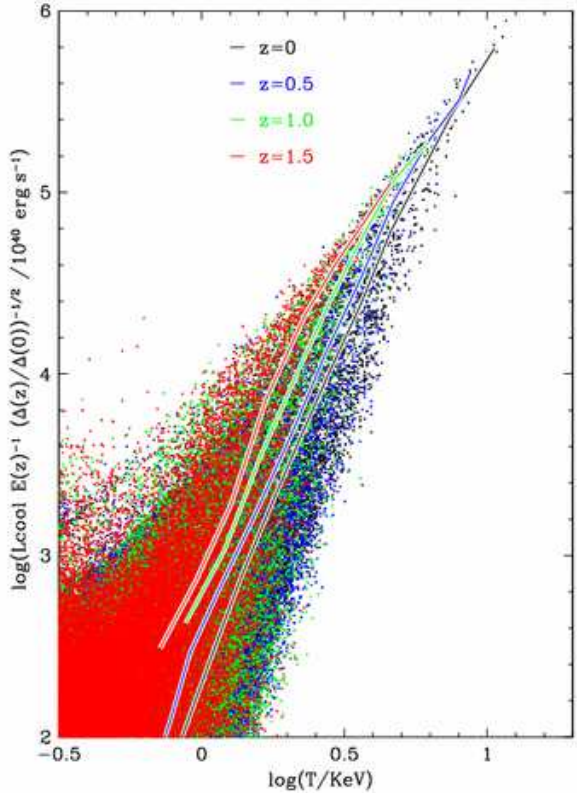}
\caption{The evolution of the L--T relation in the model. Solid lines
show the median L--T relation at different redshifts: black, $z=0$; blue,
$z=0.5$; green, $z=1.0$; red, $z=1.5$. In order to compare with the 
expected gravitational scaling of the relation, the luminosities have been
multiplied by the evolution factor, $E(z)^{-1}(\Delta(z)/\Delta(0))^{-1/2}$
defined in the text, so that the relationship will not evolve if the 
gravitational scaling dominates. Coloured points show the distribution 
of clusters and groups in the L--T at the same redshifts. Once the 
gravitational scaling has been factored out, the distribution of points 
is similar at all redshifts.}
\label{fig:lt_evo_fig}
\end{figure}

The evolution of the L--T relation is an important constraint that has not been
built into the model. The predicted evolution of the L--T relation is 
shown in Fig.~\ref{fig:lt_evo_fig}. If the development of the ICM in 
clusters were dominated by the system's gravitational collapse, the average 
density of a halo would be expected to evolve self-similarly with
redshift, tracking the density of the universe. Thus higher redshift systems
are expected to be more luminous than low redshift systems of the same temperature.
As Maughan et al.\ (2006) describe, this can be taken into account by multiplying
the observed luminosity by a factor $E(z)^{-1}(\Delta(z)/\Delta(0))^{-1/2}$,
(e.g., Bryan \& Norman 1998) where 
$E(z)=\left(\Omega_M(1+z)^3+(1-\Omega_M-\Omega_{\Lambda})(1+z)^2+\Omega_{\Lambda}\right)^{1/2}$ and
$\Delta(z)$ is well approximated by 
$18\pi^2 + 82(\Omega_M(z)-1)-39(\Omega_M(z)-1)^2$ over the range of interest
(where $\Omega_M(z) = \Omega_M(1+z)^3/E(z)$).  This reduces the luminosity 
of higher redshift clusters. Although the uncertainty is large, 
Maughan et al.\ find that this correction gives a very good description of 
the evolution seen in their data, even though the slope of the L--T relation
predicted by the gravitational scaling is inconsistent with the data. 

The evolution of the model groups and clusters is illustrated by coloured
points in Fig.~\ref{fig:lt_evo_fig}, with black, blue, green, red showing 
clusters at $z=0$, 0.5, 1.0, 1.5 respectively.  The median relations at each 
redshift are shown in colour coded lines.  As can be seen, the median
L--T relation evolves slightly faster than the expected gravitational scaling,
so that at a fixed temperature, systems are slightly brighter than would be expected
at higher redshift. A striking feature of the figure is the lack of 
higher temperature systems in the highest redshift bins, which is
due to the finite volume of the simulation and the rapid evolution of the
halo mass function. It is also notable that the scatter in the points has 
a similar dependence on temperature at all redshifts. 

At $5\keV$ the increase in predicted luminosity is a factor of 1.6 higher at $z=1.0$
than the prediction of gravitational scaling. This within the range of current 
observational constraints (Maughan et al.\ 2006). At lower temperatures, however,
the predicted evolution is stronger but harder to measure because of the large scatter 
in the expected luminosities. This is an important issue that deserves
further observational attention since the sense of the evolution predicted by 
this model is opposite to that predicted by simple preheating scenarios. In
the preheating case, the cluster luminosities are expected to scale more weakly
with redshift than the gravitational scaling because of the entropy scale
imposed on the gas becomes progressively more important as redshift increases
(see discussion in Maughan et al.\ 2006) and because cooling has had less 
time to reduce the central entropy of the system (McCarthy et al.\ 2008).

\subsection{The Properties of Galaxies and their Black Holes in the Revised Model}

\begin{figure}
  \includegraphics[width=8.6cm]{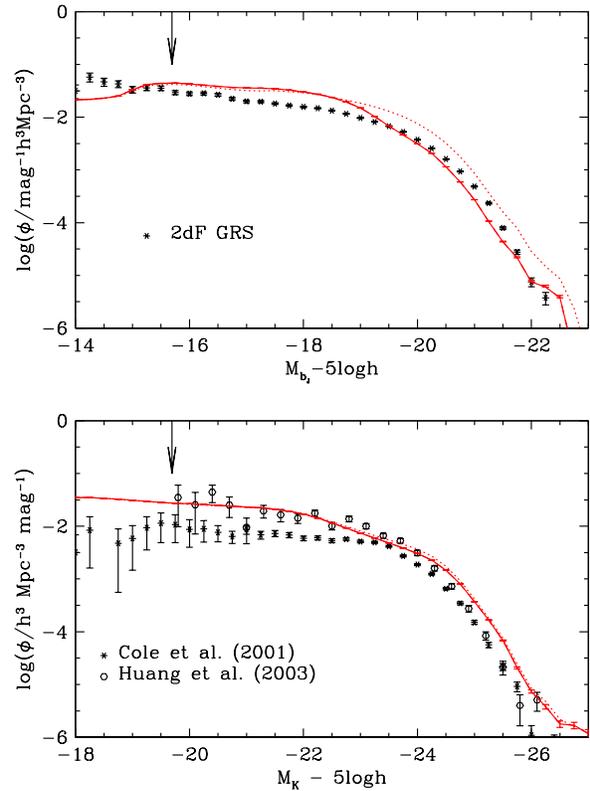}
\caption{This figure shows the $B_J$ (upper panel) and K-band (lower panel)
luminosity functions derived 
from this model. The solid line shows the model prediction, while the dotted
line shows the luminosity function obtained if no dust correction is made.
The vertical arrow shows point at which the resolution limit of the Millennium 
simulation becomes important. Black points
with error bars show recent observational measurements (see text for details). 
The model luminosity functions are broadly correct, but differ
significantly from those obtained in B06 even though the model parameters have 
been optimised for the new model.}
\label{fig:gal_lf}
\end{figure}

As we noted at the start of Section~2, including the AGN heating has
a back reaction on galaxy properties. Fig.~\ref{fig:gal_lf} illustrates the
galaxy luminosity functions that we obtain from the model. For comparison,
we show observed luminosity functions in B$_j$ from the 2dF galaxy redshift
survey by Norberg et al. (2002, upper panel) and in K-band from 
Cole et al. (2001) and Huang et al. (2003). The solid line
represents the model prediction, the dotted line shows the effect of 
removing the dust correction from the model. This figure
can be directly compared with Fig.~4 in B06. The impact of the heating of hydrostatic
haloes is very significant. If the X-ray heating is removed, the break 
in the luminosity function is far brighter. Adding the AGN heating tends to
lock haloes into the hydrostatic regime at lower masses, leading to the 
relatively good match to the luminosity function that is shown in Fig.~\ref{fig:gal_lf}.
We have also checked other galaxy properties, as described in B06. The model
reproduces the observed black hole mass -- bulge mass correlation (see below); the 
match to the colour normalisation of the blue and red sequences is 
considerably improved (as a result of the higher yield adopted --- see Font et al.\ 2008);
and the model reproduces the observed evolution of the luminosity function and
mass function at a similar level of success to B06.

While the luminosity functions have broadly the correct 
shape, normalisation and break-point, they match the observational data less well than 
the B06 model. In particular, the normalisation is rather
high, and there is a tendency for over merging to produce a tail of bright 
galaxies. In order to compensate for the latter effect, we have increased the 
merger timescale over that used in B06 (which was in turn calibrated using the 
tidal stripping calculations of Benson et al.\ 2002). This is somewhat 
unsatisfactory but may occur because the dynamical friction calculations 
of Benson et al.\ underestimate the stripping of haloes due to encounters
between satellites (in addition to the mean tidal field). 
Comparing the $B_j$ and $K$ luminosity functions reveals
an inherent tension in this model. Further fine
tuning of the model parameters cannot simultaneously match the luminosity function
in both bands. It is worth stressing that this is not because the colour
normalisation of the blue and red sequences is poor, but rather because the 
blue sequence is relatively sparsely populated at the bright end compared to B06.

The relatively poor fit to the luminosity function is not surprising given
the additional physical processes imposed on the model. Although the match is
clearly inadequate in a $\chi^2$ sense, it represents a significant step
forward over previous attempts to combine modeling of both the X-ray and 
optical properties of galaxies and clusters. The reduced goodness of fit
compared to B06, may indicate that our treatment of rapid cooling haloes  
is overly simplistic. Possibly AGN activity in these systems is able to eject some 
fraction of their baryons even before they become bound into more massive 
hydrostatic haloes. This process is beyond the scope of the present model.

\begin{figure}
  \includegraphics[width=8.6cm]{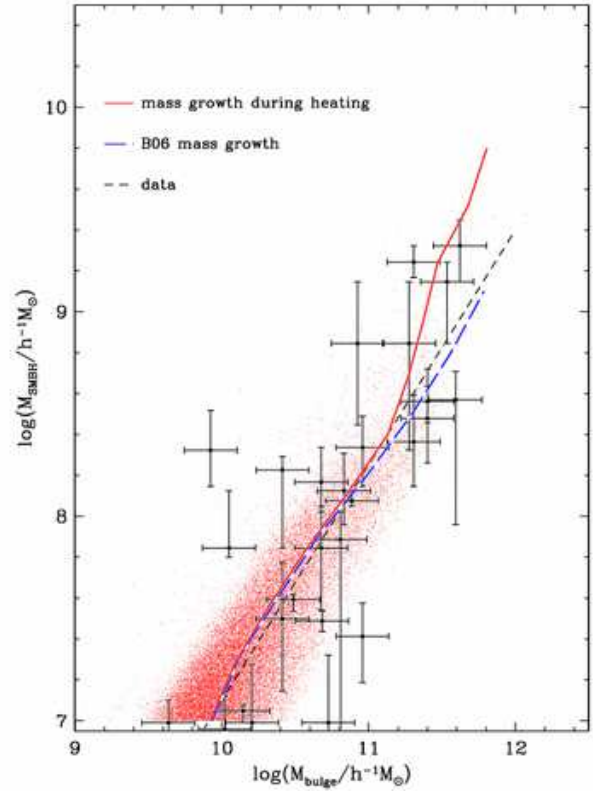}
\caption{The relation between bulge mass and black hole mass in the new 
model. A random sample of model galaxies are shown as red points, with 
the median relation plotted as a red solid line. 
Data from H{\"a}ring \& Rix 2004 is shown as large black points with error
bars, with the short dashed black line showing their best fit to the data.
The heat input from AGN in massive haloes results in a steeping of the model 
relation at large bulge masses. For comparison the median relation computed
for the new model following B06 is shown as a blue long-dashed line.}
\label{fig:bh_mbulge}
\end{figure}

The model we have presented also has implications for the black holes hosted by
galaxies in the more massive galaxies. The heating process that we have used to
eject gas from the hot haloes requires additional energy input from black holes
at the centres of hydrostatic haloes. Although this is partially
offset by the reduction in the cooling luminosity compared to the B06 model, 
the additional heating boosts black hole growth in the ``radio'' phase. 
Fig.~\ref{fig:bh_mbulge} compares the median black-hole bulge mass relation computed 
following B06 (blue line) 
with the relation derived for the new model (red line and points). The observed correlation
from H{\"a}ring \& Rix (2004) is also shown in the figure. The additional 
mass growth becomes dominant in the most massive bulges and results in a 
steeping of the relation above a bulge mass of $10^{11}\hMsol$. The most massive black 
holes exceed $10^{10}\hMsol$ in the most massive clusters. Such a steepening of the
relation is consistent with more recent analysis (Wyithe 2006) and is also
supported by analysis of the central structure of brightest cluster and group
galaxies (Lauer et al.\ 2007).

\section{Discussion and Conclusions}

At the start of this paper, our challenge was to build a model that was 
simultaneously able to account for the observed properties of 
galaxies such as the luminosity function and the X-ray properties of groups and clusters.
In this way we are seeking a model that accounts for both the number
and distribution of stars in the universe and the thermodynamic properties
of the material that is left over from the galaxy formation process. 
The importance of this second aspect is emphasised by looking at the 
mass fraction that it contains: in galaxy clusters more than 90\%
of the baryons are left over as a by-product. {In currently
popular galaxy formation models, including the B06, roughly one eighth of this
material has been processed through stars, but a much larger fraction has been 
accreted by galaxies and then ejected in the form of galactic winds.}

To rise to the challenge of modelling the ICM, we have made a relatively simple modification 
to the highly successful galaxy formation model presented in B06. In B06,
we included a ``radio mode'' of AGN feedback, allowing the energy injected
from such sources to offset radiative cooling in hydrostatic haloes. 
{On its own this model fails spectacularly to reproduce the observed
X-ray luminosities of groups and clusters}. In this paper, we have taken 
the radio-mode feedback process a step further, allowing the radio
mode feedback to eject gas from the X-ray emitting regions of hydrostatic 
haloes.

This modification of the model is largely successfully in reproducing the 
observed correlations between X-ray luminosity and system temperature. In 
particular, as well as reproducing the median slope of the relation, the 
model reproduces the large scatter in the observed luminosities of lower
temperature systems. When we take into account the distinction between the 
model virial temperatures and the observed spectroscopic temperatures the
normalisation of the relations is also in good agreement.
The steepening of the L--T relation results from gas
being ejected from the X-ray emitting regions of lower temperature groups,
and the diverse formation histories of these systems drives the 
large scatter in X-ray properties. The gas mass fractions of the model systems
agree reasonably well with the observed trends.  We also examined the 
evolution of the L--T relation predicted by the model. The observational
measurements concentrate on clusters hotter than $5\keV$. In this regime, 
the model is in agreement with current observational 
constraints. However, at lower temperatures
the model predicts groups to significantly higher luminosities (at high redshift)
than that suggested by simple gravitational scaling. The redshift dependence of the
model thus differs from that expected in a simple preheating scenario
(where we expect weaker than gravitational evolution). Although 
the large scatter in the L--T relation below $3\keV$ makes the evolution
hard to measure, this issue clearly deserves further observational effort.

{
The major problem for the model is the high level of heating required from
the AGN. This is an inevitable consequence of assuming that the heating
occurs after the system's collapse (McCarthy et al. 2008).
Thus, while the model requires only low values for the efficiency with 
which matter is accreted onto the central black hole, the predicted heating 
rates exceed the cooling rates by factors of 10 (at 5 keV) to 100 (below 1 keV)
in active systems.
This conflicts with observational estimates of cluster heating rates that
suggest that radio-mode heating is just sufficient to balance cooling in 
the more massive systems (Birzan et al. 2004; 2008; Dunn \& Fabian 2006; 2008).
However, Best et al.\ (2007) finds that the ratio heating rate increases in
lower temperature systems, and Nusser et al.\ (2006) argue that the PdV
energy is likely to underestimate the total heat input by a factor 4 to 10.
Using the revised radio luminosity calibration of Birzan et al.\ (2008),
we estimate that the model is plausibly compatible with the observed heating
rates at $T\sim 1$~keV and exceeds the observations by a factor 3 at higher 
temperatures. Clearly a much more detailed comparison of the energetics of 
the model with observational data is needed, paying careful attention to 
the observational selection effects and the difficulty in observing bubbles 
in distant or low surface brightness systems. It may also be possible that 
a more complex form for the modification of the cluster density profile 
might result in lower X-ray luminosities for a given energy input.
We have investigated whether the model can reproduce the observed L--T relation 
with lower values of the efficiency parameters $\epsilon_{SMBH}$ and $\eta_{SMBH}$.
The experiment shows that the model maintains a good match to the lowest
energy systems, but that problems occur at intermediate temperatures ($T \sim 0.5$
keV) where the L--T relation develops a pronounced break that is incompatible with 
the data.  
}

The modification of the model alters the properties of the galaxies
formed, but we find that small adjustments to the parameters in the B06 model
are able to restore reasonably good agreement with observational constraints.
In order to achieve a good fit we need to slightly raise the threshold at which 
haloes become hydrostatic and to decrease the dynamical friction orbital 
decay rate.  Without the latter modification, excessive merging tends to produce
a tail of excessively bright galaxies and a power-law (rather than exponential)
break in the luminosity function.

We have demonstrated the success of the model in reproducing the broad-brush
observational X-ray properties of groups and clusters. However, a number
of issues require closer examination and an improved model that goes
beyond a simple ad-hoc modification of the cluster density profile. In particular,
we have not attempted to address the detailed entropy profiles of these 
systems. In its current form, the model is unsuitable for this. To tackle 
such issues requires two developments. Firstly, we need to consider the
distribution of excess gas entropies and to be able to propagate this
through the merger hierarchy.  For example, McCarthy et al. (2007) show that a 
low mass substructure that has experienced little non-gravitational heating
tends to drop to the centre of a higher mass halo into which it is accreted 
with little increase in entropy. 
In this way it is quite simple to produce a small mass of low entropy 
(rapidly cooling) material
embedded in a halo of high entropy (long cooling time) gas. This complexity
is not handled by the currently model since we only track the average 
non-gravitational energy in a halo. The second aspect of the model that needs
consideration is to allow for the possibility that AGN activity and supernovae
might eject some gas from haloes even if they are not in the hydrostatic regime. This is 
a tricky issue that is difficult to assess without resort to direct numerical 
simulations (eg., Dav\'e et al. 2008). {Both of these effects will tend to amplify the energy input
at earlier epochs making the energetic demands of the model much less
daunting. The effect may be particularly relevant at intermediate temperatures
where the required heating rates appear most incompatible with the 
observational estimates.}

In summary, the current model demonstrates that we are well
on the way to understanding physical process that set the combined properties 
of the galaxy population and the thermodynamic history of the intra-cluster
medium.  While the model might not yet present a complete solution it 
provides us with great insight into the physical processes at work.

\section*{Acknowledgments}

We thank the other members of the GALFORM team (Carlos Frenk, Shaun Cole, Carlton Baugh, 
Cedric Lacey, John Helly and Rowena Malbon; see www.galform.org) for extremely helpful discussion 
and for allowing us to use the GALFORM code
as the basis of this work. RGB acknowledges the support of a Durham-University Christopherson-Knott
Fellowship.  IGMcC acknowledges support from a NSERC Postdoctoral Fellowship.
AJB acknowledges the support of the Gordon \& Betty Moore Foundation. We thank the 
referee for their thoughtful comments on the paper.

\end{document}